# Technical Analysis of Security Infrastructure in RFID Technology

**Tuhin Borgohain***
Department of Instrumentation Engineering, Assam Engineering College, Guwahati, India
Email: borgohain.tuhin@gmail.com
**Sugata Sanyal**
Corporate Technology Office, Tata Consultancy Services, Mumbai, India
Email: sugata.sanyal@tcs.com

*Corresponding Author*

**Abstract**

The paper is a technical analysis of the security infrastructure in the field of RFID technology. The paper briefly discusses the architecture of the RFID technology. Then it analyses the various features and advantages RFID technology has over the existing technologies like bar codes. This is followed by a discussion of the various disadvantages and security drawbacks of RFID technology that prevents its widespread adoption in the mainstream market. The paper concludes with a brief analysis of some of the security measures that are implemented within the RFID technology for securing up the whole infrastructure. The main aim of the paper is to focus on the drawbacks of the pre-existing security measures in RFID technology as well as to discuss the direction in which further research has to be carried out without the compromise on its unique features.

1. Introduction

Radio Frequency Identification (RFID) technology is an information exchange technology using the radio frequency for transmission of information to and fro between the various places ([1], [2], [4], [21]). Such a technology is mainly implemented in the form of tags which contain information that can be read only by specific RFID tag readers. This technology has some distinct advantages over the existing information exchange mediums like bar codes in the form of wireless interaction between the RFID tags and readers. Moreover RFID technology supports a wider array of unique IDs in comparison to bar codes ([1]) as well as has a greater data retaining capacity. And the wireless interaction between the tag and the readers facilitates automatic assessment of product with minimal human interference.

2. Overview

In section 3, we briefly discuss the architecture of the RFID tags. In section 4 we discuss the applications and advantages of RFID technology. In section 5 we discuss the disadvantages and security drawbacks of RFID technology. We discuss some security measures that can be implemented within the RFID technical framework and in general for a more secure information transmission technology in section 6. We conclude the paper in section 7.

3. RFID Architecture

The RFID technology is based on the transmission of data over electromagnetic fields (radio waves) ([4]). Here the information within the RFID tags is stored electronically. Based on the source of power derived by the RFID tags, they can be categorized mainly into two types ([25]):
    i. Active tags: These tags derive power from an external power source (In most cases, the RFID tag itself powers the active tags).
    ii. Passive tags: These tags derive energy from within them and have a limited lifetime. The passive tags power the RFID tags through electromagnetic induction.
    A RFID tag consists of the following three parts ([9]):
    i. Antenna: It performs the function of receiving energy from external power sources as well as relay information to and fro between the tag and the reader.
    ii. Semi-conductor chip attached to antenna: It co-ordinates with the antenna in the exchange of information between the tag and the reader.
    iii. Encapsulation: It acts as protective layer to the antenna and the semi-conductor chip from external environment.

A RFID tag always works in conjunction with a reader. The reader electromagnetically induces power in the tags as well as communicates with the antenna of the RFID tag.

4. Applications and Advantages of RFID Technology

Radio frequency identification technology serves as a prerequisite in IoT and it has multi-dimensional applications within the real world. The facilitation of wireless interaction between RFID tags and other electronic devices leads to ease of interaction and exchange of digital information. Some of the most prominent uses and advantages of RFID technology are as follows:

i. RFID in the transportation sector:

(a) RFID tags can be employed in car services for storing identification numbers of vehicles along with other information which can then be used for performing automatic inventories using static RFID readers appropriately placed at various locations such as parking lots, exits of garage etc. ([37], [44])

(b) In the airline industry, RFID tags can be used for efficient categorization of baggage based on routing location. This will lead to automated routing of the baggage with minimal error and a huge deduction in losses incurred due to misrouting or loss of baggage ([30], [46], [47]). However, such RFID tags can only be employed in conjunction with static RFID readers placed at conveyor belts ([9]).

(c) RFID tags can be used as re-usable tickets for public transportation systems. Such use will facilitate online purchasing or in this case online re-activation of the RFID tags for use of transportation as well as implementation of real time passenger information system ([7]).

(d) RFID tags and readers can be implemented for the operation of automated electronic tolling system ([38], [40]).

i. RFID in the health sector:

(a) RFID tags can be assigned to patients containing their health information. Moreover such an implementation along with the installation of static RFID tags in the vicinity can help in proper and timely administration of the patients' meal and other medications along with maintenance of a medical database of the patients' health progress, allergies to various foods and medicines, the various drugs administered against the respective ailments etc.

(b) RFID tags attached to medicine bottles can contain the medicinal information about the drug. Along with the use of various sensor technologies, RFID tags can also be used to determine the level of content within the bottle and appropriately facilitate alerts to users and online purchases for replenishment of content by exchange of information over the internet.

iii. RFID in hotel industry:

RFID tags in the form of smart cards are being used in the modern hotels replacing the keys to each room for enhanced security and convenience of the customers.

iv. RFID in Retail Industry and advantages of RFID over Bar Codes:

(a) RFID tags are automatically read by RFID readers in vicinity leading to reduction in manual labour and an enhanced and proactive real time tracking of products.

(b) As against bar codes, no line-of-sight is required between RFID tags and readers for reading of information ([3]).

(c) RFID tags offer higher reading rates of contained information in comparison to bar codes ([14]).

(d) RFID tags can be re-written with new information and as such are re-usable and updatable with new information about the products to which they are attached to.

(e) RFID has a much greater data capacity than a bar code.

(f) Item attached with RFID tags can be tracked in real-time point-to-point automatically thus reducing the involvement of manual labour as well as economic expenditure and time ([10]).

(g) RFID tagged items can be securely stored in the appropriate place as the tags can be programmed accordingly to send out notifications when the items leaves a specific place without proper clearance ([8]).

(h) The real time tracking facilitation by the RFID tags attached to products leads to better prediction of a particular product in the market.

(i) The RFID tags can be programmed to alert the retailers when a particular product has to be replenished in the market to prevent the "out of stock" situation as well as to prevent the loss of customers when the demand is high due to non-availability of the products.

v. RFID technology in security

(a) In some countries the passports are being attached with RFID tags for prevention of theft and forgery ([18], [28], [31], [34]).

(b) RFID technology is being rapidly implemented in the agricultural and livestock farming field for automated tracking of livestock as well as proper financial assessment against the livestock based on their presence or loss from the farmhouses ([6], [11], [22]).

(c) Being embedded with anti-cloning technology ([41], [43], [49]), the RFID cards prove to be more secure than the magnetic strips.

(d) RFID cards works when present in the vicinity of the card reader without the need of direct line of sight or direct contact like swiping.

5. Disadvantages and Security Drawbacks of RFID

The slow rate adoption of the RFID technology in the mass market even after the presence of a huge number of advantages clearly points that this technology is not yet ready for such a wide spread adoption due to a number of limiting factors. Some of the most prominent drawbacks of this technology are as follows:

i. The RFID technology proves to be unreliable in mediums like metals or liquids where RFID tags fails to be read.

ii. RFID technology, in spite of its higher reliability than electromagnetic strips and bar codes, is yet to be perfected as a significant percentage of the RFID tag fails to function properly.

iii. RFID tags can suffer from orientation problems as sometimes these tags do not interact with the readers when both are misaligned with respect to each other.

iv. The non-adoption of line of sight technology of the bar code leads to a major security drawback in parallel to its ease of use. The use of high gain antenna by a competitor can lead to loss of confidential information of products attached with RFID tags as the information contained within these tags can be easily picked up by such antennas.

v. RFID technology comes with some major financial drawback as for the use of RFID tags, a particular adopter has to first install RFID reader(s) and computer networks for assessing the information in the RFID tags. This leads to an expensive installation cost and as such a technology proves to be futile from an economic perspective.

vi. RFID technology has a signal threshold just like in the cellular communication field below which RFID technology becomes non-functioning. Moreover such a technology may malfunction in areas with high signal interference.

vii. After reading a large number of RFID tags, it may so happen in rare cases that the reader sometimes administers tag which does not exist thus leading to an error in the database entry. Such a problem is known as Ghost Tag problem in RFID.

viii. The facilitation of automated environment in the work place by the adoption of RFID tags in the various sectors of an industry like library management of assets, verification at checkouts etc. leads to decreasing demand of human workforce which leads to widespread unemployment in a region.

ix. RFID tags are susceptible to removal from an object with ease.

x. Interference between tags placed within $.125^{th}$ of an inch from each other can take place.

6. Security Measures

i. To counter the problem of Ghost Tags in RFID tags, verification technologies like CRC can be implemented in any of the three items mentioned below:

(a) RFID tag

(b) Tag reader

(c) Data from the tag.

ii. The problem arsing due to misalignment of the RFID tags with respect to the reader can be solved through the implementation of the following two solutions:

    (a) Tags having multiple axis antennas.

    (b) Multiple readers

iii. Physical locking of information in the RFID tags at the cost of rewriting capabilities of the tags will ensure proof of origins of the tags.

iv. RFID tags can be encrypted with an author's own private key. In this way an author can write information into the tag memory along with the following three security information:

    (a) Author's name

    (b) Reference to the author's public key

    (c) Algorithm used in non-encryption form.

This method helps in verifying whether the data in the tag memory has gone through any modification without the author's consent. However one of the drawbacks of this security measure is that for updating the information within the tag's memory, a key management system is required for the management of the private key.

v. RFID tags can be protected from leak of information in the presence of third party high gain antenna by enclosing the tags inside a Faraday Cage ([23], [13]) during transit.

vi. Selective active jamming of RF signals can prevent the theft of RFID tag information ([13], [15], [27]).

vii. RFID tags, if used for wireless payment, should be followed by some additional steps for verification of the user such as codes sent to cellular devices, fingerprint verification etc. This is known as multifactor authentication ([26], [33])

viii. Implementation of read detectors will help in detection of unauthorized readers in the vicinity ([24]).

ix. To avoid the misreading of tags by the RFID readers, tags can be designed to respond to frequencies set by the readers. It can also prevent leakage of information by the frequent change of frequencies by the RFID readers.

x. With advances made in the direction of integration of RFID with wireless sensor networks ([17]), it becomes imminent to implement the security measures of wireless sensor networks ([5], [12], [36], [42], [45], [48]) for securing up the RFID technology. As such implementation of sleep deprivation attack detection systems ([20]), implementation of TinySec ([19]), dynamic reconfiguration of wireless sensor network ([16]) and various other intrusion detection systems ([39]) becomes necessary for securing up the RFID technology when working in conjunction with wireless sensor networks.

## 7. Conclusion

Radio Frequency Identification technology, over its many cutting edge features and advantages, still has a long way to go for mainstream adoption due to its several drawbacks in terms of security. Adoption of security measures by compromising its re-writable feature or any of its other features brings down the whole technology to the level of the existing technologies like bar code and in the same way retaining the various features of RFID at the cost of its security makes the technology less favourable for adoption in the market. As such development of specific measures which addresses its security drawback without compromising its various features should be pushed forward for a sustainable future of this technology on the mainstream market. In addition to the development of such new security measures, the adoption of the existing data hiding techniques ([29], [32], [35]) during transmission of information can contribute to a more secure info exchange between the RFID tag and the authorized reader.